\documentclass[aps,prd,amsmath,amssymb,preprintnumbers,preprint,nofootinbib,a4paper,11pt]{revtex4-1}
\usepackage[utf8]{inputenc} 
\pdfoutput=1
\usepackage{graphicx}
\graphicspath{{./figures/}}
\usepackage{url}
\usepackage[bookmarks, pagebackref=false]{hyperref}
\usepackage[usenames,dvipsnames]{xcolor}
\definecolor{orange}{cmyk}{0,0.5,1,0}
\definecolor{rossoCP3}{cmyk}{0,.88,.77,.40}
\definecolor{graa}{rgb}{0.8,0.8,0.8}
\definecolor{blaa}{rgb}{0.2,0.2,0.6}
		\hypersetup{
			colorlinks, 
			bookmarksopen, 
			bookmarksnumbered,
			citecolor=blaa, 		
			linkcolor=rossoCP3,	
			urlcolor=rossoCP3,			
			}
\usepackage{bm}
\usepackage{bbm}
\usepackage{pxfonts}

\usepackage{placeins}
\usepackage{xspace}
\usepackage{cancel} 

\usepackage{slashed}
\usepackage[caption=false, labelformat=simple, listofformat=subsimple, labelfont=default, margin=5pt, justification=raggedright]{subfig}

	\makeatletter
		\renewcommand{\p@subfigure}{}
	\makeatother

\usepackage{natbib}

\newcommand{\ea}[1]{
\begin{align}
#1
\end{align}
}

\newcommand{\vectwo}[2]{
\left(
\begin{array}{c} #1 \\ #2 \end{array}
\right)
}

\newcommand{\beq}{\begin{eqnarray}}
\newcommand{\eeq}{\end{eqnarray}}
\newcommand{\nn}{~\nonumber \\}

\newcommand{\bmp}{\noindent\begin{minipage}{16cm}}
\newcommand{\emp}{\end{minipage}\vskip 7mm} 

\DeclareMathOperator{\tr}{Tr}

\newcommand{\mc}{\mathcal}

\newcommand{\ph}{\phantom}
\newcommand{\II}{\mathbb{I}}

\def\lsim{\mathrel{\rlap{\lower4pt\hbox{\hskip1pt$\sim$}}
    \raise1pt\hbox{$<$}}}                
\def\gsim{\mathrel{\rlap{\lower4pt\hbox{\hskip1pt$\sim$}}
    \raise1pt\hbox{$>$}}}                

\begin{document}
\newcommand{\thetitle}{Decrypting gauge-Yukawa cookbooks}
\title{\texorpdfstring{\Large\color{rossoCP3}\thetitle}{\thetitle}}
\author{Esben M\o lgaard}
\email{molgaard@cp3-origins.net}  
 \affiliation{
{\color{rossoCP3}  \rm CP}$^{\color{rossoCP3} \bf 3}${\color{rossoCP3}\rm-Origins} \& the Danish Institute for Advanced Study {\color{rossoCP3} \rm DIAS},\\ 
University of Southern Denmark, Campusvej 55, DK-5230 Odense M, Denmark.}
\begin{abstract}
For many years, theorists have calculated formulas for useful quantities in general gauge-Yukawa theories. However, these cookbooks are often very difficult to use since the general notation is far removed from practical model building. 

In this paper, we present the \emph{structure delta} which allows us to use a surprisingly convenient notation that bridges the gap between general gauge-Yukawa theories and specific models. This is particularly useful for the computation of beta functions, but can also be extended to handle spontaneous symmetry breaking, the effective potential and a variety of other quantities. We will introduce it using the standard model of particle physics and a toy model with an $\mathrm{SU}(N_c)$ gauge symmetry. 

\vspace{0.5cm}

\noindent
{ \footnotesize  \it Preprint: CP$^3$-Origins-2014-15 \& DIAS-2014-15}

\end{abstract}

\maketitle

\section{Introduction}
Much work has been done in the past to understand the general features of gauge-Yukawa theories without reference to any specific model. In particular, general formulas have been found for the beta functions and anomalous dimenions \cite{Machacek:1983tz, Machacek:1983fi, Machacek:1984zw, Pickering:2001aq, Luo:2002ti}, the effective potential \cite{Martin:2002iu}, the scalar self-energy \cite{Martin:2003it} and other useful quantities such as the presence of limit cycles in the renormalization group flow \cite{Fortin:2012hn}. These papers can be thought of as cookbooks, telling us exactly how to get the desired quantity. However, the formulas provided are expressed in a completely general notation that can be difficult to apply to a particular model, especially if the model involves many different kinds of fields, such as is the case for the standard model and its extensions.

During the course of our research \cite{Antipin:2012kc, Antipin:2013pya, Antipin:2013kia, Antipin:2013sga, Molgaard:2014mqa}, we have gradually developed a notation that allows for a transparent and generalizable translation between the notation of the general formulas and that of specific models. In the present paper, we present it in a form that should be useful to anyone interested in taking advantage of the existing formulas in the context of any particular model.

In Section \ref{sec:models}, we introduce the example models we will use to develop the notation. In Section \ref{sec:notation}, we describe the notation and show how to use it in the Yukawa and quartic sectors of the theory. We add generators of the gauge group in Section \ref{sec:generators}, and we show how to handle a situation where it is more convenient to use different fields, such as is the case after spontaneous symmetry breaking in Section \ref{sec:ssb} before concluding in Section \ref{sec:conclusion}.

\section{Example models}\label{sec:models}
Since this notation concerns how to express the same thing in different ways, it makes the most sense to introduce it through specific examples. We will in this paper consider two models; the standard model of particle physics and a toy model with fermions transforming under the fundamental and adjoint representations of the $\mathrm{SU}(N_c)$ gauge symmetry, and a large scalar sector that is a singlet under $\mathrm{SU}(N_c)$ first introduced in \cite{Antipin:2011aa}.

\subsection{The standard model}\label{sec:sm}
We write the standard model Lagrangian as follows
\ea{
  \mc L_{sm} &= \mc L_{kin} + \mc L_{mass} +\mc L_{yuk} + \mc L_{quart} \\
  \mc L_{mass} &= -\mu^2 H^\dagger H\\
  \mc L_{yuk} &= -Y^E\bar L H E - Y^D\bar Q H D - Y^U\bar Q \tilde H U + h.c.\\
  \mc L_{quart} &= -\hat\lambda (H^\dagger H)^2,
}
where $\mc L_{kin}$ contains the canonically normalized kinetic terms. The fields transform according to the representations described in table \ref{tab:smFields}, and $Y^E, Y^D$ and $Y^U$ are the Yukawa matrices of the electron, down, and up-type fields respectively. In flavor space, these are $3\times3$ matrices. Without loss of generality, it is possible to choose a basis where $Y^E$ and $Y^U$ are diagonal, and $Y^D=V\tilde Y^D$ where $\tilde Y^D$ is diagonal and $V$ is the unitary CKM-matrix.

\begin{table}[h!]
  \[ \begin{array}{c|c c c|c} \hline \hline
    {\rm Fields} & \left[ SU(3)_c \right] &  \left[ SU(2)_W \right] &  \left[ U(1)_Y \right] & \text{Chirality} \\ \hline 
     L & 1 & 2 & -\frac12 & L\\
     E & 1 & 1 & -1 & R\\
     Q & 3 & 2 & \ph-\frac16 & L\\
     D & 3 & 1 & -\frac13 & R \\
     U & 3 & 1 & \ph-\frac23 & R\\
    \hline
     H & 1 & 2 & \ph-\frac12 & \\
       \hline \hline \end{array}%
  \]%
\caption{Transformation properties of the standard model fields under the three constituent gauge groups.}%
\label{tab:smFields}%
\end{table}

\subsection{$\mathrm{SU}(N_c)$ toy model}\label{sec:ams}
We will also consider the model given by the Lagrangian \eqref{eq:Lams} first introduced in \cite{Antipin:2011aa} and futher developed in \cite{Antipin:2012kc,Antipin:2013pya,Antipin:2013kia}. It has an $\mathrm{SU}(N_c)$ gauge symmetry under which there are fermions transforming in the fundamental, conjugate fundamental and adjoint representations, and also features a global $\mathrm{SU}(N_f)_L\times\mathrm{SU}(N_f)_R$ symmetry  (see table \ref{tab:amsFields} for the details). The fundamental and conjugate fundamental Weyl fermions can be thought of as forming a Dirac vector fermion in analogy with QCD, but we find the present description more convenient.
\ea{
\mathcal{L}&= 
- \frac{1}{4} F^{\mu \nu}F_{\mu \nu} +i \lambda  \slashed{D} \bar\lambda+ i\bar q \slashed{D} q + i\bar{\tilde{q}} \slashed{D} \tilde q + \partial_\mu H ^\dagger \partial^\mu H + (y_H\tilde q H q +h.c.) - u_1 (\tr [H ^\dagger H])^2 -u_2\tr[(H ^\dagger H )^2] \ ,
 \label{eq:Lams}
}
where $\lambda$ is the adjoint fermion, $q$ and $\tilde q$ are (anti)fundamental fermions and $H$ is a singlet scalar field. 


\begin{table}[h!]
   \[
   \begin{array}{c|c|c c c c|c}
     \hline
     \hline
     Fields & [SU(N_c)] & SU(N_f)_L & SU(N_f)_R & U(1)_V & U(1)_{AF} & \text{Chirality}\\\hline 
     \lambda & Adj &$1$&$1$&$0$ & $1$ & L\\
     q &  \Box  & \overline{\Box} & 1 & \tfrac{N_f-N_c}{N_c} & -\tfrac{N_c}{N_f} & L\\ 
     \tilde{q} & \overline{\Box}  & 1 & \Box & -\tfrac{N_f-N_c}{N_c} & -\tfrac{N_c}{N_f} & L \\
     \hline
     H & 1 & \Box & \overline{\Box} & 0 & \tfrac{2N_c}{N_f} & \\
     G_{\mu} & Adj & 1 & 1 & 0 & 0 & \\ \hline \hline
   \end{array}
   \]
  \caption{The field content of the model and the related symmetries}
  \label{tab:amsFields}
\end{table}

We will consider this model when introducing spontaneous symmetry breaking in Section \ref{sec:ssb}.

\section{Lagrangian notation}\label{sec:notation}

The general formulas presented in various papers (see for example \cite{Machacek:1983tz, Machacek:1983fi,Machacek:1984zw,Pickering:2001aq}) all assume a basic Lagrangian of the form
\ea{
  \mathcal L &= \mathcal L_{kin} - \frac{1}{2}\left(y_{JK;A}\Psi_J\Psi_K\Phi_A + h.c.\right) - \frac{1}{4!}\lambda_{ABCD}\Phi_A\Phi_B\Phi_C\Phi_D \label{genL},
}
where $\Psi_J$ is an all-encompassing fermion field transforming under some (in general) reducible representation of the gauge group, and we are (and will be in the following) summing over repeated indices. Without loss of generality, we may assume that it has a definite chirality, and that all of its component fields are Weyl spinors. Similarly, $\Phi_A$ is an all-encompassing scalar field transforming under a reducible representation of the gauge group. Scalar or fermion mass terms, and scalar cubic terms can be included by introducing a non-propagating dummy real scalar field $\Phi_{\hat D}$ \cite{Luo:2002ti}. The relevant operators can then be expressed as
\ea{
  \mathcal L_1 &= - \frac{1}{2}\left(y_{JK;\hat D}\Psi_J\Psi_K\Phi_{\hat D} + h.c.\right) - \frac{1}{4!}\lambda_{AB\hat D\hat D}\Phi_A\Phi_B\Phi_{\hat D}\Phi_{\hat D}- \frac{1}{4!}\lambda_{ABC\hat D}\Phi_A\Phi_B\Phi_C\Phi_{\hat D} \label{genL-rel},
}
where $y_{JK;\hat D}\Phi_{\hat D}=(m_f)_{JK}$, $\lambda_{AB\hat D\hat D}\Phi_{\hat D}\Phi_{\hat D} = 2m_{AB}^2$ and $\lambda_{ABC\hat D}\Phi_{\hat D} = h_{ABC}$. The beta functions for the relevant operators can then be obtained from those of the corresponding marginal operators.

Any gauge-Yukawa theory can, in principle be put on this form. How to actually accomplish this in a general way is not at all obvious, and it is the focus of the present paper.

To accomplish this task, we introduce an object which organizes the specific fields (such as $\bar L$, $E$ and $H$ in the standard model) within the abstract fields $\Psi_J$ and $\Phi_A$, and keeps track of overall indices as well as specific field indices. We call this the \emph{structure delta}
\ea{
  \Delta^{S}_{J;\{j\}} \label{strucDelta}
}
where $S$ takes values from the names of the fields in the theory ($\bar L$, $E$ and $H$); $J$ is the overall fermion index from \eqref{genL}; and $\{j\}$ covers the gauge and flavor indices of the field $S$.

This new symbol obeys the following summation rule
\ea{
  \Delta^{S}_{J;\{j\}}\Delta^{S'}_{J;\{j'\}} &= \delta^{SS'}\prod_{\{j\},\{j'\}}\delta_{jj'},
}
and has the fundamental property that $\Psi_J\Delta^{S}_{J;\{j\}}=S_{\{j\}}$ (or $\Phi_J\Delta^{S}_{J;\{j\}}=S_{\{j\}}$ if $S$ is a scalar field). In the specific case of the standard model lepton sector, this is realized in the following manner
\ea{
  \Psi_J\Delta^{\bar L;g_2,f_L}_{J}=\bar L^{g_2,f_L}, \quad \Psi_J\Delta^{E}_{J;f_E}=E_{f_E} \quad \mathrm{and}\quad \Phi_A\Delta^{H}_{A;g_2,c}=H_{g_2,c}. \label{strucDeltaSM}
}

\subsection{Yukawa interactions}
We first illustrate how to express general Yukawa interactions by considering the leptonic part of the standard model Yukawa interaction with all flavor and gauge indices written explicitly. It is
\ea{
  \mathcal L_{Yuk,Lep} = (Y^E)^{f_E}_{f_L}\bar L^{g_2,f_L} H_{g_2} E_{f_E} + h.c.\ ,
}
where $g_2$ is the SU(2) gauge index, and $f_L$ and $f_E$ are the flavor indices of the lepton doublet and electron-like singlet respectively. There is a subtlety here because the standard model Higgs is a complex scalar, and thus have twice as many degrees of freedom as its gauge index would suggest. We take this into account by adding a complex index $c$ and introducing the symbol $o^c$ with the property that $o^1 = 1$ and $o^2 = i$. Then
\ea{
  \mathcal L_{Yuk,Lep} = (Y^E)^{f_E}_{f_L}o^c\bar L^{g_2,f_L} H_{g_2,c}E_{f_E} + h.c.\ . \label{lepL}
}

Using the structure delta \eqref{strucDelta}, we can now write the lepton Yukawa Lagrangian \eqref{lepL} as
\ea{
  \mathcal L_{Yuk,Lep} &= \frac{1}{2}(Y^E)^{f_E}_{f_L}o^c\left(\Psi_J\Delta^{\bar L;g_2,f_L}_{J}\Psi_K\Delta^{E}_{K;f_E}+\Psi_K\Delta^{\bar L;g_2,f_L}_{K}\Psi_J\Delta^{E}_{J;f_E}\right)\Phi_A\Delta^{H}_{A;g_2,c}  + h.c.\\
&=\frac{1}{2}(Y^E)^{f_E}_{f_L}o^c\left(\Delta^{\bar L;g_2,f_L}_{J}\Delta^{E}_{K;f_E}+\Delta^{\bar L;g_2,f_L}_{K}\Delta^{E}_{J;f_E}\right)\Delta^{H}_{A;g_2,c}\Psi_J\Psi_K\Phi_A  + h.c.\ ,
}
and simply read off
\ea{
  y_{JK;A}^{(Lep)} = (Y^E)^{f_E}_{f_L}o^c\left(\Delta^{\bar L;g_2,f_L}_{J}\Delta^{E}_{K;f_E}+\Delta^{\bar L;g_2,f_L}_{K}\Delta^{E}_{J;f_E}\right)\Delta^{H}_{A;g_2,c}.
}

We can now use an equivalent procedure to find $y_{JK;A}^{(Up)}$ and $y_{JK;A}^{(Down)}$, and then construct the Yukawa coupling for the entire standard model is just the sum,
\ea{
  y_{JK;A} = y_{JK;A}^{(Lep)} + y_{JK;A}^{(Up)} + y_{JK;A}^{(Down)} \ .
}

\subsection{Quartic interactions}
We can find the quartic coupling $\lambda_{ABCD}$ in an analogous way,
\ea{
  \mathcal L_{quart} &= \hat\lambda\left(H^\dagger H\right)^2.
}
$H$ can be written as a complex vector,
\ea{
  H &= \frac{1}{\sqrt{2}}\vectwo{H_{1,1} + i H_{1,2}}{H_{2,1} + i H_{2,2}},
} 
thus
\ea{
  H^\dagger H &= \frac{1}{2}(H_{1,1} - i H_{1,2})(H_{1,1} + i H_{1,2}) + \frac{1}{2}(H_{2,1} - i H_{2,2})(H_{2,1} + i H_{2,2})\\
 &= \frac{1}{2}(H_{1,1}^2 + H_{1,2}^2 + H_{2,1}^2 + H_{2,2}^2),
}
which we can write as 
\ea{
  H^\dagger H &= \frac{1}{2}H_{g_2,c}H_{g_2,c}\ .
}
Thus, in terms of the structure deltas, 
\ea{
  \mathcal L_{quart,H} &= \frac{\hat\lambda}{4} H_{g_2,c}H_{g_2,c}H_{g'_2,c'}H_{g'_2,c'}\\
  &= \frac{\hat\lambda}{4}\frac{1}{4!}\sum_{perms}\Delta^{H}_{A;g_2,c}\Delta^{H}_{B;g_2,c}\Delta^{H}_{C;g'_2,c'}\Delta^{H}_{D;g'_2,c'}\Phi_A\Phi_B\Phi_C\Phi_D,
}
where ${\sum_{perms}}$ is a sum over all possible permutation of $A,B,C,D$, and the factor of $\frac{1}{4!}$ enters to compensate for the sum.

We can now read off
\ea{
  \lambda_{ABCD} &= \frac{\hat\lambda}{4}\sum_{perms}\Delta^{H}_{A;g_2,c}\Delta^{H}_{B;g_2,c}\Delta^{H}_{C;g'_2,c'}\Delta^{H}_{D;g'_2,c'}
}

\section{Generators}\label{sec:generators}
Following the notation of \cite{Pickering:2001aq}, we refer to the generator of the (reducible) scalar representation as $S^{\alpha}_{AB}$, where $A$ and $B$ are general scalar indices, and $\alpha$ is the group index running from 1 to $d(G)$. In the case of a semi-simple group, this is generalized to $S^{t;\alpha}_{AB}$ where $t$ labels the simple subgroup. Equivalently, the generator of the spinor representation is  given as $R^{\alpha}_{JK}$, and $R^{t;\alpha}_{JK}$ if the group is semi-simple.

The normalization of $S$ and $R$ is such that 
\ea{
  \tr(S^\alpha S^\beta) &= \delta^{\alpha\beta}T(S), & S^{\alpha}_{AC}S^{\alpha}_{CB} &= C(S)_{AB},\\
  \tr(R^\alpha R^\beta) &= \delta^{\alpha\beta}T(R), & R^{\alpha}_{JL}R^{\alpha}_{LK} &= C(R)_{JK},\\
  f^{\alpha\gamma\delta}f^{\beta\gamma\delta} &= \delta^{\alpha\beta}C(G), & \delta^{\alpha\alpha} &= d(G),
}
where $T(\cdot)$ is the Dynkin index of the representation, $C(\cdot)$ is the quadratic Casimir of the representation, with $C(G)$ in particular being the quadratic Casimir of the adjoint, and $d(G)$ is the dimension of the group.

We can find $S^{\alpha}_{AB}$ by summing over the generators and structure deltas of each scalar species. For $\mathrm{U}(1)$ this is particularly simple as the generator is just the charge of the field. Since we are decomposing the complex scalars into their real components, each set of indices corresponds to a single real scalar field. Furthermore, since the generators must be hermitian, this implies that they must be imaginary and anti-symmetric. The expression for the $\mathrm{U}(1)$ generator is thus
\ea{
  S^{\alpha}_{AB} = \sum_S-i Q_S \epsilon^{c}{}_{c'}\delta_{f_S'}^{f_S}\Delta^{S}_{A;c,f_S}\Delta^{S;c',f_S'}_{B},
}
where $Q_S$ is the charge of the scalar species $S$. The sign is conventional, and has been chosen such that the charges of the standard model fields are as listed in Table \ref{tab:smFields}.

For a non-abelian group, things are more complicated as the fields now transform in non-trivial representations of the group. Since each scalar field under consideration is still real, we must still ensure that each term in the generator is imaginary and anti-symmetric under an exchange of \emph{all} of the indices associated with the field. 


We first observe that since the generators, $T^{\alpha;a}{}_{b}$, only carry gauge indices, the flavor indices are contracted through a delta function. Secondly, the generators are either real and symmetric, or imaginary and anti-symmetric. To ensure that the final expression is imaginary and anti-symmetric, we must multiply by $i\epsilon^{c}{}_{c'}$  in the former case, in complete analogy with the $\mathrm{U}(1)$ case above, but in the latter we must instead multiply by $\delta_{c'}^{c}$. We show that the following construction is imaginary and anti-symmetric in either case:
\ea{
  &T^{\alpha;a}{}_{b}(\delta_{c'}^{c}+i\epsilon^{c}{}_{c'})-T^{\alpha;a}_{b}(\delta_{c'}^{c}-i\epsilon^{c}{}_{c'})\\
  \underset{\text{\ph{antisymmetric}}}{=}{}&T^{\alpha;a}{}_{b}\delta_{c'}^{c}+iT^{\alpha;a}{}_{b}\epsilon^{c}{}_{c'}-T^{\alpha;a}_{b}\delta_{c'}^{c}+iT^{\alpha;a}_{b}\epsilon^{c}{}_{c'}\\
  \underset{\text{\ph{an}symmetric\ph{ti}}}{=}{}&2iT^{\alpha;a}{}_{b}\epsilon^{c}{}_{c'}\\
  \underset{\text{antisymmetric}}{=}{}&2T^{\alpha;a}{}_{b}\delta_{c'}^{c}\ ,
}
where $T^{\alpha;a}_{b}$ is the transpose of $T^{\alpha;a}{}_{b}$.

The final expression for the generator of the scalar representation is thus:
\ea{
  S^{\alpha}_{AB} = \sum_S-\frac{1}{2}\left(T_S^{\alpha;a}{}_{b}(\delta_{c'}^{c}+i\epsilon^{c}{}_{c'})-T_{S;b}^{\alpha}{}^{a}(\delta_{c'}^{c}-i\epsilon^{c}{}_{c'})\right)\delta_{f_S'}^{f_S}\Delta^{S}_{A;c,f_S,a}\Delta^{S;c',f_S',b}_{B}
}
where $T_S^{\alpha;a}{}_{b}$ is the generator of the representation under which the scalar species $S$ transform. $a$ and $b$ are the gauge indices of the representation, $\alpha$ is the gauge index of the group, and $f_S,f_S'$ are the flavor indices of the scalar species $S$. The sign is again conventional and a change of sign here will correspond to a change in the sign of the structure constants of the group.

The fermion case is equivalent to the scalar case, but simpler since we do not need to keep track of the complex index $c$. In the abelian case, we have
\ea{
  R^{\alpha}_{JK} = \sum_SQ_S\delta_{f_S'}^{f_S}\Delta^{S}_{J;f_S}\Delta^{S;f_S'}_{K}.
}
and in the non-abelian
\ea{
  R^{\alpha}_{JK} = \sum_ST_S^{\alpha;k}{}_{j}\delta_{f_S'}^{f_S}\Delta^{S}_{J;f_S,j}\Delta^{S;f_S',b}_{K}
} 

The expressions for the generators can be generalized to a semi-simple group by including an index $t$ for the subgroup in question and a product $\prod_{t'\neq t}\delta^{a_t}_{b_t}$ over the other subgroups.

\subsection{Generator rewritings}
When working with this notation in the special case of $\mathrm{SU}(N)$ groups, it can be convenient to use the relation \cite{MacFarlane:1968vc}
\ea{
  f^{\alpha\beta\gamma} &= -2 i (T^\alpha_{ij}T^\beta_{jk}-T^\beta_{ij}T^\alpha_{jk})T^\gamma_{ki}
}
where $T^\alpha_{ij}$ is the generator of the fundamental representation of the group. The reason this is convenient is that it makes it easy to apply the Fierz identity
\ea{
  T^\alpha_{ij}T^\alpha_{kl} = \frac{1}{2}\left(\delta_{il}\delta_{jk}-\frac{1}{N}\delta_{ij}\delta_{kl}\right),
}
which holds for the generators of the fundamental representation of $\mathrm{SU}(N)$.

Similarly, if one considers a model where the matter fields transform in a higher-dimensional representation of the group (such as the Georgi-Glashow $\mathrm{SU}(5)$ GUT \cite{Georgi:1974sy}, or models of walking technicolor \cite{Dietrich:2005jn}), it is useful to express the generators of the two index symmetric or antisymmetric representation in terms of the generators of the fundamental representation. To do this, we simply write
\ea{
  \hat T^{\alpha}_{\hat j\hat k} = \hat T^{\alpha}_{\{j_1j_2\}\{k_1k_2\}} = \frac12\left(\delta_{j_2k_2}T^\alpha_{j_1k_1}\pm\delta_{j_2k_1}T^\alpha_{j_1k_2}\pm\delta_{j_1k_2}T^\alpha_{j_2k_1}+\delta_{j_1k_1}T^\alpha_{j_2k_2}\right),
}
where $\hat T^{\alpha}_{\hat j\hat k}$ is the generator of the (anti)symmetric representation and $\{j_1j_2\}$ refers to the (anti)symmetrized indices $(j_1j_2)$ \big($[j_1j_2]$\big). 

It is easy to check that this expansion reproduces the well known results for the Dynkin index, quadratic casimir and dimension of the two index (anti)symmetric representation,
\ea{
  \hat T^{\alpha}_{\hat j\hat k}\hat T^{\beta}_{\hat k\hat j} &=\frac14\big(\delta_{j_2k_2}T^\alpha_{j_1k_1}\pm\delta_{j_2k_1}T^\alpha_{j_1k_2}\pm\delta_{j_1k_2}T^\alpha_{j_2k_1}+\delta_{j_1k_1}T^\alpha_{j_2k_2}\big)\big(\delta_{k_2j_2}T^\beta_{k_1j_1}\pm\delta_{k_2j_1}T^\beta_{k_1j_2}\pm\delta_{k_1j_2}T^\beta_{k_2j_1}+\delta_{k_1j_1}T^\beta_{k_2j_2}\big)\nn
  &= \frac{N\pm2}{2}\delta^{\alpha\beta}\\
  \hat T^{\alpha}_{\hat j\hat k}\hat T^{\alpha}_{\hat k\hat l} &=\frac14\big(\delta_{j_2k_2}T^\alpha_{j_1k_1}\pm\delta_{j_2k_1}T^\alpha_{j_1k_2}\pm\delta_{j_1k_2}T^\alpha_{j_2k_1}+\delta_{j_1k_1}T^\alpha_{j_2k_2}\big)\big(\delta_{k_2l_2}T^\alpha_{k_1l_1}\pm\delta_{k_2l_1}T^\alpha_{k_1l_2}\pm\delta_{k_1l_2}T^\alpha_{k_2l_1}+\delta_{k_1l_1}T^\alpha_{k_2l_2}\big)\nn
  &= \frac{(N\pm2)(N\mp1)}{2N}(\delta_{j_1l_1}\delta_{j_2l_2}\pm\delta_{j_1l_2}\delta_{j_2l_1}) = \frac{(N\pm2)(N\mp1)}{N}\delta_{\hat j\hat l}\\
  \delta_{\hat j\hat j} &= \frac{N}{(N\pm2)(N\mp1)}\hat T^{\alpha}_{\hat j\hat k}\hat T^{\alpha}_{\hat k\hat j} = \frac{N(N\pm1)}{2}
}

\section{Spontaneous symmetry breaking}\label{sec:ssb}
To describe the situation where one desires to express the model under consideration using a different set of fields at different scales, such as is the case when chiral symmetry breaking happens, we consider the model we introduced in Section \ref{sec:ams}. Before chiral symmetry breaking, the scalar sector is expressed as a general $N_f\times N_f$ complex matrix $H$, but afterwards, this matrix develops a vacuum expectation values and has the parametrization (see \cite{Antipin:2011aa} for details) 
\ea{
  H_{l}^r = \frac{v + \phi + i \eta}{\sqrt{2N_f}}\delta_{l}^r+h^a T^{a,r}_{l}+i\pi^aT^{a,r}_{l}
}
where $v$ is the vacuum expectation value, $\phi$ and $h^a$ gain masses, $\eta$ and $\pi^a$ are the Goldstone bosons, and $T^{a,r}_{l}$ are the generators of the fundamental representation of $SU(N_f)$.

In the notation established in Section \ref{sec:notation}, this corresponds to exchanging the structure delta $\Delta^H_{A;lr}$ for a sum over the new structure deltas. In general, this is done as follows
\ea{
  \Delta^S_{A;\{a\}} \to \sum_{i}\Omega^{S_i}_{\{a\};\{a_i\}}\Delta^{S_i}_{A;\{a_i\}}\ ,
}
where the index $i$ specifies which of the descendant fields that particular term describes, and $\Omega$ contains the relevant information on the gauge and flavor structure of the breaking.

In the particular case mentioned above, the breaking pattern is
\ea{
  \Delta^{H;r}_{A;l} \to \frac{v}{\sqrt{2N_f}}\delta_{l}^r\II_A+\frac{1}{\sqrt{2N_f}}\delta_{l}^r\Delta^\phi_{A}+\frac{i}{\sqrt{2N_f}}\delta_{l}^r\Delta^\eta_{A}+T^{a,r}_{l}\Delta^h_{A;a}+iT^{a,r}_{l}\Delta^\pi_{A;a}\label{AMSbreaking}
}
where we have introduced a new object $\II_A$ which is defined by the action 
\ea{
  \II_A\Phi_A=1 && \II_J\Psi_J=1.
}
This has the useful property that it can turn Yukawa and quartic terms into mass terms. To illustrate this, let us consider the Yukawa coupling of \eqref{eq:Lams} in the notation introduced previously in this paper
\ea{
  y_H\tilde{q}^g_r H_l^r q^l_{g} = \frac{y_H}{2}\Big(\Delta^{\tilde{q};g}_{J;r} \Delta^{q;l}_{K;g}+\Delta^{\tilde{q};g}_{K;r} \Delta^{q;l}_{J;g}\Big)\Delta^{H;r}_{A;l}\Psi_J\Psi_K\Phi_A \ .
}
Under the breaking pattern \eqref{AMSbreaking}, this becomes
\ea{
  y_H\tilde{q}^g_r H_l^r q^l_{g} \to{}& \frac{y_H}{2}\Big(\Delta^{\tilde{q};g}_{J;r} \Delta^{q;l}_{K;g}+\Delta^{\tilde{q};g}_{K;r} \Delta^{q;l}_{J;g}\Big)\Bigg(\frac{v\II_A}{\sqrt{2N_f}}\delta_{l}^r+\frac{\Delta^\phi_{A}+i\Delta^\eta_{A}}{\sqrt{2N_f}}\delta_{l}^r+T^{a,r}_{l}\Delta^h_{A;a}+iT^{a,r}_{l}\Delta^\pi_{A;a}\Bigg)\Psi_J\Psi_K\Phi_A\\
={}& \frac{y_Hv}{2\sqrt{2N_f}}\Big(\Delta^{\tilde{q};g}_{J;l} \Delta^{q;l}_{K;g}+\Delta^{\tilde{q};g}_{K;l} \Delta^{q;l}_{J;g}\Big)\Psi_J\Psi_K\nn
& +\frac{y_H}{2}\Big(\Delta^{\tilde{q};g}_{J;r} \Delta^{q;l}_{K;g}+\Delta^{\tilde{q};g}_{K;r} \Delta^{q;l}_{J;g}\Big)\Bigg(\frac{\Delta^\phi_{A}+i\Delta^\eta_{A}}{\sqrt{2N_f}}\delta_{l}^r+T^{a,r}_{l}\Delta^h_{A;a}+iT^{a,r}_{l}\Delta^\pi_{A;a}\Bigg)\Psi_J\Psi_K\Phi_A\ ,
}
and we see that the first term is in fact a fermion mass term since it contains a dimension 1 operator and 2 fermion fields.

A completely analogous analysis can be performed for the quartic sector after breaking, and this leads to scalar mass terms and cubic interactions, as well as new quartic terms.

\section{Concluding remarks}\label{sec:conclusion}
We have presented a new notation which makes it simple to write the Lagrangian of a specific gauge-Yukawa theory in a manner that allows the direct application of formulas derived for general gauge-Yukawa theories. This enables theorists and model builders to quickly get expressions for the beta function, effective potential etc. for specific theories under examination.

\section*{Acknowledgements}
The author would like to thank the C.N. Yang Institute for Theoretical Physics at the State University of New York at Stony Brook for its hospitality while part of this work was done; and Jens Krog, Oleg Antipin and Matin Mojaza for careful reading and fruitful discussion about the manuscript. The CP$^3$-Origins centre is partially funded by the Danish National Research Foundation, grant number DNRF90.

\bibliographystyle{apsrev4-1}
\bibliography{notationBib,esbenMolgaard}

\begin{thebibliography}{17}%
\makeatletter
\providecommand \@ifxundefined [1]{%
 \@ifx{#1\undefined}
}%
\providecommand \@ifnum [1]{%
 \ifnum #1\expandafter \@firstoftwo
 \else \expandafter \@secondoftwo
 \fi
}%
\providecommand \@ifx [1]{%
 \ifx #1\expandafter \@firstoftwo
 \else \expandafter \@secondoftwo
 \fi
}%
\providecommand \natexlab [1]{#1}%
\providecommand \enquote  [1]{``#1''}%
\providecommand \bibnamefont  [1]{#1}%
\providecommand \bibfnamefont [1]{#1}%
\providecommand \citenamefont [1]{#1}%
\providecommand \href@noop [0]{\@secondoftwo}%
\providecommand \href [0]{\begingroup \@sanitize@url \@href}%
\providecommand \@href[1]{\@@startlink{#1}\@@href}%
\providecommand \@@href[1]{\endgroup#1\@@endlink}%
\providecommand \@sanitize@url [0]{\catcode `\\12\catcode `\$12\catcode
  `\&12\catcode `\#12\catcode `\^12\catcode `\_12\catcode `\%12\relax}%
\providecommand \@@startlink[1]{}%
\providecommand \@@endlink[0]{}%
\providecommand \url  [0]{\begingroup\@sanitize@url \@url }%
\providecommand \@url [1]{\endgroup\@href {#1}{\urlprefix }}%
\providecommand \urlprefix  [0]{URL }%
\providecommand \Eprint [0]{\href }%
\providecommand \doibase [0]{http://dx.doi.org/}%
\providecommand \selectlanguage [0]{\@gobble}%
\providecommand \bibinfo  [0]{\@secondoftwo}%
\providecommand \bibfield  [0]{\@secondoftwo}%
\providecommand \translation [1]{[#1]}%
\providecommand \BibitemOpen [0]{}%
\providecommand \bibitemStop [0]{}%
\providecommand \bibitemNoStop [0]{.\EOS\space}%
\providecommand \EOS [0]{\spacefactor3000\relax}%
\providecommand \BibitemShut  [1]{\csname bibitem#1\endcsname}%
\let\auto@bib@innerbib\@empty
\bibitem [{\citenamefont {Machacek}\ and\ \citenamefont
  {Vaughn}(1983)}]{Machacek:1983tz}%
  \BibitemOpen
  \bibfield  {author} {\bibinfo {author} {\bibfnamefont {M.~E.}\ \bibnamefont
  {Machacek}}\ and\ \bibinfo {author} {\bibfnamefont {M.~T.}\ \bibnamefont
  {Vaughn}},\ }\href {\doibase 10.1016/0550-3213(83)90610-7} {\bibfield
  {journal} {\bibinfo  {journal} {Nucl.Phys.}\ }\textbf {\bibinfo {volume}
  {B222}},\ \bibinfo {pages} {83} (\bibinfo {year} {1983})}\BibitemShut
  {NoStop}%
\bibitem [{\citenamefont {Machacek}\ and\ \citenamefont
  {Vaughn}(1984)}]{Machacek:1983fi}%
  \BibitemOpen
  \bibfield  {author} {\bibinfo {author} {\bibfnamefont {M.~E.}\ \bibnamefont
  {Machacek}}\ and\ \bibinfo {author} {\bibfnamefont {M.~T.}\ \bibnamefont
  {Vaughn}},\ }\href {\doibase 10.1016/0550-3213(84)90533-9} {\bibfield
  {journal} {\bibinfo  {journal} {Nucl.Phys.}\ }\textbf {\bibinfo {volume}
  {B236}},\ \bibinfo {pages} {221} (\bibinfo {year} {1984})}\BibitemShut
  {NoStop}%
\bibitem [{\citenamefont {Machacek}\ and\ \citenamefont
  {Vaughn}(1985)}]{Machacek:1984zw}%
  \BibitemOpen
  \bibfield  {author} {\bibinfo {author} {\bibfnamefont {M.~E.}\ \bibnamefont
  {Machacek}}\ and\ \bibinfo {author} {\bibfnamefont {M.~T.}\ \bibnamefont
  {Vaughn}},\ }\href {\doibase 10.1016/0550-3213(85)90040-9} {\bibfield
  {journal} {\bibinfo  {journal} {Nucl.Phys.}\ }\textbf {\bibinfo {volume}
  {B249}},\ \bibinfo {pages} {70} (\bibinfo {year} {1985})}\BibitemShut
  {NoStop}%
\bibitem [{\citenamefont {Pickering}\ \emph {et~al.}(2001)\citenamefont
  {Pickering}, \citenamefont {Gracey},\ and\ \citenamefont
  {Jones}}]{Pickering:2001aq}%
  \BibitemOpen
  \bibfield  {author} {\bibinfo {author} {\bibfnamefont {A.}~\bibnamefont
  {Pickering}}, \bibinfo {author} {\bibfnamefont {J.}~\bibnamefont {Gracey}}, \
  and\ \bibinfo {author} {\bibfnamefont {D.}~\bibnamefont {Jones}},\ }\href
  {\doibase 10.1016/S0370-2693(01)00624-4} {\bibfield  {journal} {\bibinfo
  {journal} {Phys.Lett.}\ }\textbf {\bibinfo {volume} {B510}},\ \bibinfo
  {pages} {347} (\bibinfo {year} {2001})},\ \Eprint
  {http://arxiv.org/abs/hep-ph/0104247} {arXiv:hep-ph/0104247 [hep-ph]}
  \BibitemShut {NoStop}%
\bibitem [{\citenamefont {Luo}\ \emph {et~al.}(2003)\citenamefont {Luo},
  \citenamefont {Wang},\ and\ \citenamefont {Xiao}}]{Luo:2002ti}%
  \BibitemOpen
  \bibfield  {author} {\bibinfo {author} {\bibfnamefont {M.-x.}\ \bibnamefont
  {Luo}}, \bibinfo {author} {\bibfnamefont {H.-w.}\ \bibnamefont {Wang}}, \
  and\ \bibinfo {author} {\bibfnamefont {Y.}~\bibnamefont {Xiao}},\ }\href
  {\doibase 10.1103/PhysRevD.67.065019} {\bibfield  {journal} {\bibinfo
  {journal} {Phys.Rev.}\ }\textbf {\bibinfo {volume} {D67}},\ \bibinfo {pages}
  {065019} (\bibinfo {year} {2003})},\ \Eprint
  {http://arxiv.org/abs/hep-ph/0211440} {arXiv:hep-ph/0211440 [hep-ph]}
  \BibitemShut {NoStop}%
\bibitem [{\citenamefont {Martin}(2002)}]{Martin:2002iu}%
  \BibitemOpen
  \bibfield  {author} {\bibinfo {author} {\bibfnamefont {S.~P.}\ \bibnamefont
  {Martin}},\ }\href {\doibase 10.1103/PhysRevD.66.096001} {\bibfield
  {journal} {\bibinfo  {journal} {Phys.Rev.}\ }\textbf {\bibinfo {volume}
  {D66}},\ \bibinfo {pages} {096001} (\bibinfo {year} {2002})},\ \Eprint
  {http://arxiv.org/abs/hep-ph/0206136} {arXiv:hep-ph/0206136 [hep-ph]}
  \BibitemShut {NoStop}%
\bibitem [{\citenamefont {Martin}(2004)}]{Martin:2003it}%
  \BibitemOpen
  \bibfield  {author} {\bibinfo {author} {\bibfnamefont {S.~P.}\ \bibnamefont
  {Martin}},\ }\href {\doibase 10.1103/PhysRevD.70.016005} {\bibfield
  {journal} {\bibinfo  {journal} {Phys.Rev.}\ }\textbf {\bibinfo {volume}
  {D70}},\ \bibinfo {pages} {016005} (\bibinfo {year} {2004})},\ \Eprint
  {http://arxiv.org/abs/hep-ph/0312092} {arXiv:hep-ph/0312092 [hep-ph]}
  \BibitemShut {NoStop}%
\bibitem [{\citenamefont {Fortin}\ \emph {et~al.}(2013)\citenamefont {Fortin},
  \citenamefont {Grinstein},\ and\ \citenamefont {Stergiou}}]{Fortin:2012hn}%
  \BibitemOpen
  \bibfield  {author} {\bibinfo {author} {\bibfnamefont {J.-F.}\ \bibnamefont
  {Fortin}}, \bibinfo {author} {\bibfnamefont {B.}~\bibnamefont {Grinstein}}, \
  and\ \bibinfo {author} {\bibfnamefont {A.}~\bibnamefont {Stergiou}},\ }\href
  {\doibase 10.1007/JHEP01(2013)184} {\bibfield  {journal} {\bibinfo  {journal}
  {JHEP}\ }\textbf {\bibinfo {volume} {1301}},\ \bibinfo {pages} {184}
  (\bibinfo {year} {2013})},\ \Eprint {http://arxiv.org/abs/1208.3674}
  {arXiv:1208.3674 [hep-th]} \BibitemShut {NoStop}%
\bibitem [{\citenamefont {Antipin}\ \emph
  {et~al.}(2012{\natexlab{a}})\citenamefont {Antipin}, \citenamefont
  {Di~Chiara}, \citenamefont {Mojaza}, \citenamefont {Mølgaard},\ and\
  \citenamefont {Sannino}}]{Antipin:2012kc}%
  \BibitemOpen
  \bibfield  {author} {\bibinfo {author} {\bibfnamefont {O.}~\bibnamefont
  {Antipin}}, \bibinfo {author} {\bibfnamefont {S.}~\bibnamefont {Di~Chiara}},
  \bibinfo {author} {\bibfnamefont {M.}~\bibnamefont {Mojaza}}, \bibinfo
  {author} {\bibfnamefont {E.}~\bibnamefont {Mølgaard}}, \ and\ \bibinfo
  {author} {\bibfnamefont {F.}~\bibnamefont {Sannino}},\ }\href {\doibase
  10.1103/PhysRevD.86.085009} {\bibfield  {journal} {\bibinfo  {journal}
  {Phys.Rev.}\ }\textbf {\bibinfo {volume} {D86}},\ \bibinfo {pages} {085009}
  (\bibinfo {year} {2012}{\natexlab{a}})},\ \Eprint
  {http://arxiv.org/abs/1205.6157} {arXiv:1205.6157 [hep-ph]} \BibitemShut
  {NoStop}%
\bibitem [{\citenamefont {Antipin}\ \emph
  {et~al.}(2013{\natexlab{a}})\citenamefont {Antipin}, \citenamefont {Gillioz},
  \citenamefont {Mølgaard},\ and\ \citenamefont {Sannino}}]{Antipin:2013pya}%
  \BibitemOpen
  \bibfield  {author} {\bibinfo {author} {\bibfnamefont {O.}~\bibnamefont
  {Antipin}}, \bibinfo {author} {\bibfnamefont {M.}~\bibnamefont {Gillioz}},
  \bibinfo {author} {\bibfnamefont {E.}~\bibnamefont {Mølgaard}}, \ and\
  \bibinfo {author} {\bibfnamefont {F.}~\bibnamefont {Sannino}},\ }\href
  {\doibase 10.1103/PhysRevD.87.125017} {\bibfield  {journal} {\bibinfo
  {journal} {Phys.Rev.}\ }\textbf {\bibinfo {volume} {D87}},\ \bibinfo {pages}
  {125017} (\bibinfo {year} {2013}{\natexlab{a}})},\ \Eprint
  {http://arxiv.org/abs/1303.1525} {arXiv:1303.1525 [hep-th]} \BibitemShut
  {NoStop}%
\bibitem [{\citenamefont {Antipin}\ \emph
  {et~al.}(2013{\natexlab{b}})\citenamefont {Antipin}, \citenamefont {Krog},
  \citenamefont {Mølgaard},\ and\ \citenamefont {Sannino}}]{Antipin:2013kia}%
  \BibitemOpen
  \bibfield  {author} {\bibinfo {author} {\bibfnamefont {O.}~\bibnamefont
  {Antipin}}, \bibinfo {author} {\bibfnamefont {J.}~\bibnamefont {Krog}},
  \bibinfo {author} {\bibfnamefont {E.}~\bibnamefont {Mølgaard}}, \ and\
  \bibinfo {author} {\bibfnamefont {F.}~\bibnamefont {Sannino}},\ }\href
  {\doibase 10.1007/JHEP09(2013)122} {\bibfield  {journal} {\bibinfo  {journal}
  {JHEP}\ }\textbf {\bibinfo {volume} {1309}},\ \bibinfo {pages} {122}
  (\bibinfo {year} {2013}{\natexlab{b}})},\ \Eprint
  {http://arxiv.org/abs/1303.7213} {arXiv:1303.7213 [hep-ph]} \BibitemShut
  {NoStop}%
\bibitem [{\citenamefont {Antipin}\ \emph
  {et~al.}(2013{\natexlab{c}})\citenamefont {Antipin}, \citenamefont {Gillioz},
  \citenamefont {Krog}, \citenamefont {Mølgaard},\ and\ \citenamefont
  {Sannino}}]{Antipin:2013sga}%
  \BibitemOpen
  \bibfield  {author} {\bibinfo {author} {\bibfnamefont {O.}~\bibnamefont
  {Antipin}}, \bibinfo {author} {\bibfnamefont {M.}~\bibnamefont {Gillioz}},
  \bibinfo {author} {\bibfnamefont {J.}~\bibnamefont {Krog}}, \bibinfo {author}
  {\bibfnamefont {E.}~\bibnamefont {Mølgaard}}, \ and\ \bibinfo {author}
  {\bibfnamefont {F.}~\bibnamefont {Sannino}},\ }\href {\doibase
  10.1007/JHEP08(2013)034} {\bibfield  {journal} {\bibinfo  {journal} {JHEP}\
  }\textbf {\bibinfo {volume} {1308}},\ \bibinfo {pages} {034} (\bibinfo {year}
  {2013}{\natexlab{c}})},\ \Eprint {http://arxiv.org/abs/1306.3234}
  {arXiv:1306.3234 [hep-ph]} \BibitemShut {NoStop}%
\bibitem [{\citenamefont {Mølgaard}\ and\ \citenamefont
  {Shrock}(2014)}]{Molgaard:2014mqa}%
  \BibitemOpen
  \bibfield  {author} {\bibinfo {author} {\bibfnamefont {E.}~\bibnamefont
  {Mølgaard}}\ and\ \bibinfo {author} {\bibfnamefont {R.}~\bibnamefont
  {Shrock}},\ }\href@noop {} {\bibfield  {journal} {\bibinfo  {journal}
  {Phys.Rev.}\ }\textbf {\bibinfo {volume} {D}},\ \bibinfo {pages} {accepted
  for publication} (\bibinfo {year} {2014})},\ \Eprint
  {http://arxiv.org/abs/1403.3058} {arXiv:1403.3058 [hep-th]} \BibitemShut
  {NoStop}%
\bibitem [{\citenamefont {Antipin}\ \emph
  {et~al.}(2012{\natexlab{b}})\citenamefont {Antipin}, \citenamefont {Mojaza},\
  and\ \citenamefont {Sannino}}]{Antipin:2011aa}%
  \BibitemOpen
  \bibfield  {author} {\bibinfo {author} {\bibfnamefont {O.}~\bibnamefont
  {Antipin}}, \bibinfo {author} {\bibfnamefont {M.}~\bibnamefont {Mojaza}}, \
  and\ \bibinfo {author} {\bibfnamefont {F.}~\bibnamefont {Sannino}},\ }\href
  {\doibase 10.1016/j.physletb.2012.04.050} {\bibfield  {journal} {\bibinfo
  {journal} {Phys.Lett.}\ }\textbf {\bibinfo {volume} {B712}},\ \bibinfo
  {pages} {119} (\bibinfo {year} {2012}{\natexlab{b}})},\ \Eprint
  {http://arxiv.org/abs/1107.2932} {arXiv:1107.2932 [hep-ph]} \BibitemShut
  {NoStop}%
\bibitem [{\citenamefont {MacFarlane}\ \emph {et~al.}(1968)\citenamefont
  {MacFarlane}, \citenamefont {Sudbery},\ and\ \citenamefont
  {Weisz}}]{MacFarlane:1968vc}%
  \BibitemOpen
  \bibfield  {author} {\bibinfo {author} {\bibfnamefont {A.}~\bibnamefont
  {MacFarlane}}, \bibinfo {author} {\bibfnamefont {A.}~\bibnamefont {Sudbery}},
  \ and\ \bibinfo {author} {\bibfnamefont {P.}~\bibnamefont {Weisz}},\ }\href
  {\doibase 10.1007/BF01654302} {\bibfield  {journal} {\bibinfo  {journal}
  {Commun.Math.Phys.}\ }\textbf {\bibinfo {volume} {11}},\ \bibinfo {pages}
  {77} (\bibinfo {year} {1968})}\BibitemShut {NoStop}%
\bibitem [{\citenamefont {Georgi}\ and\ \citenamefont
  {Glashow}(1974)}]{Georgi:1974sy}%
  \BibitemOpen
  \bibfield  {author} {\bibinfo {author} {\bibfnamefont {H.}~\bibnamefont
  {Georgi}}\ and\ \bibinfo {author} {\bibfnamefont {S.}~\bibnamefont
  {Glashow}},\ }\href {\doibase 10.1103/PhysRevLett.32.438} {\bibfield
  {journal} {\bibinfo  {journal} {Phys.Rev.Lett.}\ }\textbf {\bibinfo {volume}
  {32}},\ \bibinfo {pages} {438} (\bibinfo {year} {1974})}\BibitemShut
  {NoStop}%
\bibitem [{\citenamefont {Dietrich}\ \emph {et~al.}(2005)\citenamefont
  {Dietrich}, \citenamefont {Sannino},\ and\ \citenamefont
  {Tuominen}}]{Dietrich:2005jn}%
  \BibitemOpen
  \bibfield  {author} {\bibinfo {author} {\bibfnamefont {D.~D.}\ \bibnamefont
  {Dietrich}}, \bibinfo {author} {\bibfnamefont {F.}~\bibnamefont {Sannino}}, \
  and\ \bibinfo {author} {\bibfnamefont {K.}~\bibnamefont {Tuominen}},\ }\href
  {\doibase 10.1103/PhysRevD.72.055001} {\bibfield  {journal} {\bibinfo
  {journal} {Phys.Rev.}\ }\textbf {\bibinfo {volume} {D72}},\ \bibinfo {pages}
  {055001} (\bibinfo {year} {2005})},\ \Eprint
  {http://arxiv.org/abs/hep-ph/0505059} {arXiv:hep-ph/0505059 [hep-ph]}
  \BibitemShut {NoStop}%
\end{thebibliography}%

\end{document}